\documentclass[preprint,pre,aps]{revtex4}
\usepackage{graphicx}
\usepackage{epstopdf}
\usepackage{amsmath}
\usepackage{amsfonts}
\begin{document}
\title{Periodic ordering of clusters in a one-dimensional lattice model.}
\author{J. P\c ekalski,  A. Ciach}
\affiliation{Institute of Physical Chemistry,
 Polish Academy of Sciences, 01-224 Warszawa, Poland}
\author{N. G. Almarza}
\affiliation{Instituto de Qu{\'\i}mica F{\'\i}sica Rocasolano, CSIC, Serrano 119, E-28006 Madrid, Spain }
\begin{abstract}
A generic lattice model for systems containing particles interacting with short-range attraction
 long-range repulsion (SALR) potential that can be solved exactly in one dimension is introduced.
We assume attraction $J_1$ between the first neighbors and repulsion $J_2$ between the third neighbors.
 The ground state of the model shows existence of two homogeneous phases (gas and liquid) for $J_2/J_1<1/3$.  In addition 
to the homogeneous phases, 
 the third phase with periodically distributed clusters appears for $J_2/J_1>1/3$. 
Phase diagrams obtained in the self-consistent mean-field approximation for a range of values of $J_2/J_1$ show very 
rich behavior, including reentrant melting, and coexistence of two periodic phases (one with  strong and the other 
one with  weak order) terminated at a critical point. 
We present exact solutions  for the equation of state as well as for the  correlation 
function for characteristic values of  $J_2/J_1$. 
 Based on the exact results, for  $J_2/J_1>1/3$
we predict pseudo-phase transitions to the ordered cluster phase indicated by a rapid change of density for a very narrow
 range of pressure, and by a very large correlation length for thermodynamic states where the periodic phase is stable in mean
 field. For $1/9<J_2/J_1<1/3$ the correlation function decays monotonically below certain temperature, 
whereas above this temperature
exponentially damped oscillatory behavior is obtained. Thus, even though macroscopic phase separation is energetically 
favored and appears for weak repulsion at $T=0$, local spatial inhomogeneities appear for finite $T$. 
Monte Carlo simulations in canonical ensemble show that specific heat has a maximum for low density $\rho$ that we associate with 
formation of living clusters, and if the repulsion is strong, another maximum for $\rho = 1/2$.

\end{abstract}
\maketitle

\section{Introduction}

Most of biologically relevant macromolecules, or particles in soft-matter systems are charged and repel each other with
 screened electrostatic forces \cite{israel:11:0,barrat:03:0,stradner:04:0,campbell:05:0}. On the other hand, 
 complex solvents in biological or soft-matter systems may induce effective attraction
 between the macromolecules. Important examples of the effective attraction include depletion forces resulting from the
 presence of small objects like nonadsorbing polymers \cite{dijkstra:99:0,campbell:05:0,buzzaccaro:10:0}, solvophobic 
 attraction \cite{iglesias:12:0,shukla:08:0} or thermodynamic 
Casimir forces \cite{veatch:07:0,hertlein:08:0,gambassi:09:0,machta:12:0} 
resulting from critical adsorption near a critical demixing point in a binary (or  multicomponent) solvent. The sum of
 all interactions has often the form of short-range attraction long-range repulsion (SALR) potential
 \cite{sear:99:0,pini:00:0,imperio:04:0,imperio:07:0,imperio:06:0,pini:06:0,archer:07:0,ciach:08:1,archer:08:0,archer:08:1,archer:07:1,ciach:10:1,schwanzer:10:0,candia:06:0,pini:06:0,pini:00:0,toledano:09:0}. 

Despite of the importance of the SALR potential for biological and soft-matter systems only preliminary and incomplete 
results for phase diagrams have been obtained so far 
\cite{imperio:04:0,imperio:07:0,imperio:06:0,ciach:08:1,archer:08:0,archer:08:1,archer:07:0,archer:07:1,ciach:10:1,schwanzer:10:0,candia:06:0,pini:06:0,pini:00:0}. 
It is clear that the phase diagrams depend on the strength of the
 repulsion, and for strong enough repulsion can be completely different than in simple fluids. In the latter systems the
 phase behavior is determined by the competition between the entropy favoring the disordered phase and the energy
 favoring formation of a spherical droplet of the dense phase. Volume fraction of the particles determines only the 
size of this droplet. In the case of the SALR potential there is additional competition between the energy and the 
chemical potential (or the volume fraction) of particles. The SALR potential favors  formation of spherical clusters whose size
 is determined by the range of the attraction, and the distance between the clusters is determined by the range of 
the repulsion. The structure minimizing the energy is possible only for sufficiently small volume fraction, however. 
 The increase of the chemical potential leads to transitions from spherical to cylindrical clusters, next to a network,
 then to layers of particles, and finally to inverse structures (occupied regions replaced by voids)
 \cite{archer:07:1,ciach:08:1,archer:08:0,candia:06:0,campbell:05:0,ciach:10:1}. Regions of stability of the disordered 
fluid  and different ordered phases, 
as well as location and order of phase-transitions between them for various shapes of the SALR potential are only partially known.

The difficulty in determination of phase diagrams in experimental studies results from the presence of many metastable 
phases and large characteristic time scales, by which instantaneous rather than average states have been
  observed \cite{stradner:04:0,campbell:05:0}. The question of the average structure in thermodynamic 
equilibrium may be clarified in future confocal microscopy dynamical studies.  
In simulations it is much more difficult to study inhomogeneous structures with some kind of periodic ordering and
 moreover the results are restricted to the chosen form of interactions
 \cite{imperio:04:0,imperio:07:0,imperio:06:0,archer:07:1,candia:06:0,toledano:09:0}. 
In theoretical studies the results relay on 
approximations, usually of the mean-field (MF) type 
\cite{ciach:08:1,archer:07:0,archer:07:1,archer:08:0,archer:08:1,ciach:10:1,pini:00:0,pini:06:0,schwanzer:10:0}. 
Validity of  MF for the SALR type potentials was only partially verified \cite{archer:07:1,ciach:08:1},
 and it is not clear which predictions of the MF 
approximation and to what extent are valid on a qualitative or on a semiquantitative level. Preliminary results within  
field-theoretic approach show that fluctuations may lead to substantial modification of the equation of state 
\cite{ciach:12:0} and  to a fluctuation-induced first order phase transition \cite{ciach:08:1,ciach:11:0}, as in 
 magnetic systems with competing interactions \cite{brazovskii:75:0,bak:76:0}. We should also mention that the accurate
 liquid theories such as SCOZA \cite{dickman:96:0} are unfortunatelly limited to uniform phases, and only a lack of
 solutions is an indicator
 of possible phases with periodic ordering on the mesoscopic length scale \cite{pini:00:0,pini:06:0,archer:07:0}. 
 Further studies including 
comparison between MF type results and simulations, as in Ref. \cite{archer:07:1} are necessary. 
It would be also desirable to 
introduce a simple model that could be solved exactly. Exact results could serve for testing  general theoretical 
predictions and for verification of the range of validity of approximate theories. No simple model that could play
the same role as the two dimensional Ising model played for 
simple systems was proposed so far.

An exactly solvable two-dimensional model for the SALR potential cannot be proposed at the moment. 
To fill this gap at least partially, 
we introduce in this work a one-dimensional (1d) lattice model with attraction $J_1$ between the first and repulsion $J_2$ 
between the third neighbors. 
 Exact solutions can be obtained  for the equation of state (EOS) as well as for the density and the  correlation function for
 the whole range of  $J_2/J_1$.
There are no phase transitions in 1d models with finite range of interactions, but pseudo-phase transitions may
 exist for low temperatures. 
Moreover, it is interesting to 
compare the exact and MF results.  It is also of interest to compare thermodynamics and
 structure of the disordered phase in simple fluid and in the SALR-potential systems.
 The results can give interesting information about pretransitional ordering in the latter systems. 
Effects of the precursors of the ordered structures on structure, compressibility and specific heat were already studied in
 approximate theories and in simulations
 \cite{pini:00:0,pini:06:0,schwanzer:10:0,archer:07:0,imperio:04:0,imperio:06:0,imperio:07:0}. Based on the exact
 results for the correlation function and for the EOS, we can verify if the repulsion can lead to
 significantly increased
 compressibility and to oscillatory decay of correlations on a mesoscopic length scale, as found in Ref.\cite{pini:06:0,archer:07:0,ciach:08:1,archer:08:0,archer:08:1,archer:07:1}. 

 In MF phase transitions are present even in one dimension, therefore MF in 1d systems 
can shed light on some general features of phase diagrams in two or three dimensions. 

Formation of living (or dynamic) clusters may be associated with large energy fluctuations,
 especially when the clusters easily form or dissociate. The specific heat in the SALR system was studied in
 Ref.\cite{imperio:06:0}. For fixed densities the maximum of the specific heat at some temperature was interpreted as
 a signature of the
 transition between the homogeneous and inhomogeneous fluids. To calculate the specific heat
in our model we perform Monte Carlo (MC) simulations in the canonical ensemble. We calculate the specific heat for several
 temperatures as  function of density for strong and for weak repulsion. 

The one dimensional model considered here, apart from yielding general information on periodic ordering in the SALR-potential
 systems, describes several physical systems that are interesting by their own. In the first place it can represent
 charged particles in a  presence of depletant at a three-phase coexistence line, or adsorbed at nanotubes or microtubules.
 Another example is a linear backbone polymer with monomers containing sites  binding  particles or ions that attract or 
repel each other when bound to first or third neighbors on the backbone respectively.  Our model can answer the question of
 spontaneous  formation of ordered periodic patterns on linear substrates.

In the next section the model is introduced and the ground state is determined and discussed. 
Sec.3 is devoted to the MF approximation. Exact results and MC simulations are described in secs.4 and 5 respectively.
Sec.6 contains a short summary and discussion. 
\section{The model and its ground state}
\subsection{The model}
We consider an open system in equilibrium with a reservoir with temperature $T$ and chemical potential $\mu_p$. The
 interaction $h$ between
 the particles and  the  nanotubes, microtubules  or binding sites
 plays analogous role as the chemical potential, and we introduce $\mu=\mu_p+h$.
The particles can occupy lattice sites labeled by $x$ taking integer values, $1\le x\le L$ and we assume periodic
 boundary conditions (PBC), i.e. $L+1\equiv 1, 0\equiv L$. Each microstate is described by  
$\{\hat\rho(x)\}\equiv(\hat\rho(1),...,\hat\rho(L))$, where the occupancy operator $\hat\rho(x)=1$ or $\hat\rho(x)=0$ 
when the site $x$ is occupied or empty respectively. The probability of the  microstate $\{\hat\rho(x)\}$ is given by 
\begin{equation}
p[\{\hat\rho(x)\}]=\frac{e^{-\beta H[\{\hat\rho(x)\}]}}{\Xi},
\end{equation}  
where $\Xi$ is the normalization constant, and $\beta=(k_BT)^{-1}$ with $k_B$ denoting the Boltzmann constant.
We have introduced for convenience the thermodynamic Hamiltonian containing the energy and the chemical potential term,
\begin{equation}
\label{H}
  H [\{\hat\rho\}] = \frac{1}{2} \sum_{x=1}^L\sum_{x'=1}^L \hat{\rho}( x) V ( x-x')\hat{\rho}(x') 
- \mu \sum_{x=1}^L \hat{\rho}( x),
  \end{equation}
 where  the interaction potential is
\begin{equation}
 V(\Delta{ x}) = -J_1 \Big( \delta(\Delta{ x} + 1) + \delta(\Delta{ x} - 1)\Big)
+ J_2 \Big( \delta(\Delta{ x} + 3) + \delta(\Delta{ x} - 3)\Big).
\label{nasz_potencjal}
\end{equation}

The Hamiltonian (\ref{H}) can be rewritten in terms of the 
``unoccupancy'' operator $\hat \nu (x)=1-\hat\rho(x)$ ($\hat\nu(x)=1,0$ for an empty and full site $x$ respectively)
\begin{equation}
\label{Hnu}
  H [\{\hat\nu\}] = \frac{1}{2} \sum_{x=1}^L\sum_{x'=1}^L \hat{\nu}( x) V ( x-x')\hat{\nu}(x') 
+ (\mu-V_0) \sum_{x=1}^L \hat{\nu}( x) +L\Big(\frac{V_0}{2}-\mu\Big)
  \end{equation}
where $V_0=\sum_x V(x)=2(J_2-J_1)$. Note that  Eq.(\ref{H}) in terms of $\hat\rho$ and Eq.(\ref{Hnu}) in terms
 of $\hat\nu$  have the same form for $\mu=V_0/2$. Moreover, the probability of the microstate
 $\{\hat\rho( x) \}$ for $\mu=V_0/2-\Delta\mu$ is the same as the probability of the ``negative'' of this  microstate, 
$\{1-\hat\rho( x) \}$, for $\mu=V_0/2+\Delta\mu$. Because of the above particle-hole symmetry
 the phase diagrams must be symmetric 
with respect to the symmetry axis $\mu=J_2-J_1$.

We choose $J_1$ as the energy unit and introduce dimensionless variables for any quantity $X$ with 
dimension of energy as $X^*=X/J_1$, in particular
\begin{equation}
\label{dimensionless}
T^*=k_BT/J_1,J^*=J_2/J_1,\mu^*=\mu/J_1.
 \end{equation}
\subsection{The ground state}

\begin{figure}[th]
\includegraphics[scale=1]{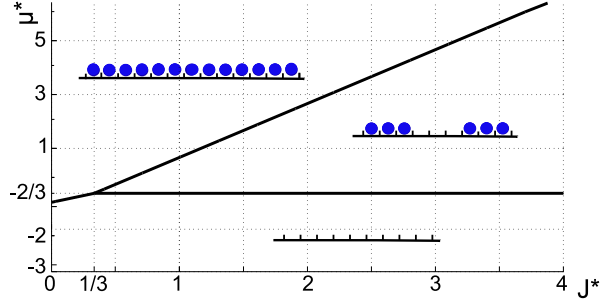}
\caption{ Ground state of the considered model. The repulsion to attraction ratio $J^*$ and the chemical potential $\mu^*$ 
are dimensionless (see (\ref{dimensionless})). The coexistence lines are $\mu^{*}_{gas-cond} = -1 + J^* $, $
\mu^{*}_{gas-per} =-2/3$ and $
\mu^{*}_{cond-per} =-4/3 + 2J^*$. 
 Schematic illustration of the three phases is shown in the insets inside the region of stability of each phase.}
\label{ground_state}
\end{figure}

The grand potential 
\begin{equation}
\Omega=-pL=-k_BT\ln \Xi=U-TS-\mu N 
 \end{equation}
where $p, U, S,N$ are pressure, internal energy, entropy and  average number of particles respectively,
 reduces to the minimum of $H [\{\hat\rho(x)\}]$ for $T=0$ and fixed $L$. In the case of periodic phases the bulk 
properties must be determined for $L=ln$, where $l$ is the period of density oscillations and $n$ is integer. 
We consider  $\omega^*=-p^*=H^* [\{\hat\rho(x)\}]/(ln)$ for two homogeneous phases, one with all sites empty (gas) and the other one with all sites occupied (liquid) and for a periodic phase where three occupied neighboring sites are followed by $l-3$ empty sites with $l\ge 6$. For these phases we have
\begin{displaymath}
\omega^*= \left\{ \begin{array}{ll}
0 & \textrm{ empty lattice (gas)}\\
-\frac{2+3\mu^*}{l} & \textrm{periodic, $l\ge 6$}\\
J^*-1-\mu^*  & \textrm{ full occupancy (liquid)}\\
\end{array} \right.
\end{displaymath} 
Two phases can coexist for thermodynamic states such that $\omega^*$ in these phases takes the same value.
 The  $(J^*,\mu^*)$ phase diagram for $T^*=0$ is shown in Fig.\ref{ground_state}.

Note that for $\mu^*=-2/3$ the $\omega^*$ of the periodic phase 
is independent of $l$ if $l\ge 6$. This is because when in the empty lattice 3 neighboring 
cells become occupied, the associated change of $H^*$ is $-2-3\mu^*$. Because the interaction range is 3, for $\mu^*=-2/3$
 a  triple of occupied cells can be
separated  from another triple of occupied cells by $l-3$ empty cells for any $l\ge 6$.  Such a state can be interpreted
 as a cluster fluid
 that can be stable, however, for a single value of the chemical potential, $\mu^*=-2/3$.  When $2+3\mu^*>0$, the lowest 
value of $\omega^*$ corresponds to  $l=6$, and the periodic phase  with period 6 is stable.  The gas and periodic
 phases coexist for $\mu^*=-2/3$. Because of this coexistence, arbitrary separation between the clusters,
 whose number is also arbitrary (but smaller than $L/6$),
 can be interpreted as  arbitrarily small droplets (larger than $6$ in the case of the periodic phase) of these phases. 
As a result, an arbitrary number of interfaces can be formed. This is possible when the surface tension between the gas and the
 periodic phases vanishes.

Similarly, creation of a triple of empty sites in the fully occupied lattice leads to the change of $H^*$ which is 
$\Delta H^*=-6J^*+4+3\mu^*$. At the coexistence between the fully occupied lattice and the periodic phase $6J^*-4-3\mu^*=0$, 
hence the separation between the three empty neighboring cells (bubbles) can be arbitrary (but $\ge 3$). Again, such a state
 can be interpreted as a  fluid of bubbles, or as a coexistence between the liquid and periodic phases in the case of 
vanishing surface tension.
Note the similarity between this property of our model and the very low surface tension between water- or oil- rich
 phases and microemulsion. At $T^*=0$ formation of the microemulsion is associated with vanishing surface tension in the
lattice model for the water-oil-surfactant mixture\cite{ciach:89:0}.

\section{MF approximation}
\subsection{Short background
}
In the self-consistent MF approximation the Hamiltonian (\ref{H}) is approximated by
\begin{equation}
\label{HMF}
 H^{MF}[\{\hat\rho(x)\}]=\sum_{ x} \big[-(h(x)+\mu )\hat\rho(x)+\frac{1}{2} h(x)\bar\rho(x)\Big]
\end{equation}
where the mean-field acting on the site $x$ is 
\begin{equation}
 \label{h}
h(x)=-\sum_{x'}V(x-x')\bar\rho(x')
\end{equation}
and the MF average density satisfies the self-consistent equation
\begin{equation}
\label{ror}
 \bar\rho(x)=\frac{e^{\beta(h(
x)+\mu)}}{1+e^{\beta(h(x)+\mu)}}.
\end{equation}
The grand statistical sum
\begin{equation}
\label{XiMF}
 \Xi=\prod_x \Big[
e^{-\frac{\beta}{2}h(x)\bar\rho(x)}\Big(1+e^{\beta(h(
x)+\mu)}\Big)
\Big]
\end{equation}
together with (\ref{ror}) after some algebra leads to the grand potential of the form \cite{ciach:11:1}
\begin{eqnarray}
\label{wzomeg}
 \Omega &=&\sum_{ x_1=1}^L\sum_{x_2=1}^L\Bigg\{\frac{1}{2}\bar\rho(
x_1)\bar\rho(x_2) 
V(x_1-x_2)\\
\nonumber
&+&\delta^{Kr}(x_1-x_2)
f_h( \bar\rho(x_1))
\Bigg\}-\mu\sum_{x=1}^L\bar\rho(x).
\end{eqnarray}
where in the lattice models 
\begin{eqnarray}
 f_h( \rho)=-k_BTs( \rho)=k_BT
\Big[ \rho\ln ( \rho)+(1- \rho)\ln(1-
 \rho)
\Big]
\label{fh}.
\end{eqnarray}
Local minima of (\ref{wzomeg}) satisfy Eq.(\ref{ror}) (see Ref.\cite{ciach:11:1}).

Eq.(\ref{ror}) can be solved by iterations for different initial conditions. Stability regions of different phases 
and first-order transitions between them can be obtained by comparing $\omega=\Omega/(ln)$ for different forms of $\bar\rho(x)$.
 In practice systems with the size $l$ and PBC represent one period of the phases with the period $l$, and we have considered 
$6\le l\le 50$.  

\subsection{Stability analysis}
Boundary of stability of the disordered phase 
can be found by analyzing the
 second derivative of $\beta\Omega$ with respect to the density profile $\bar\rho(x)$. The disordered fluid is stable as
 long as this derivative, 
\begin{equation}
\label{C}
C(x,x')=\frac{\partial \beta\Omega}{\partial \bar\rho(x)\partial \bar\rho(x')},
\end{equation}
 is positive definite.  The  disordered phase is at the boundary of stability when the smallest eigenvalue of (\ref{C})
 vanishes for $\rho(x)=const$. For interactions depending only on $x-x'$ the  quadratic part of $\beta\Omega$ (bilinear form) is diagonal  in 
Fourier representation, and the eigenvalues of $C$ 
are given by
\begin{equation}
 \tilde C(k)=\beta^*\tilde V(k)+\frac{1}{\bar\rho(1-\bar\rho)},
\end{equation}
where
\begin{equation}
\label{V(k)def}
\tilde{V}(k) = \sum_{x} V^{*}(x) e^{ikx}  .
\end{equation}
In this model 
\begin{equation}
\label{V(k)}
 \tilde{V}(k) = - 2 \cos k + 2 J^{*} \cos 3k.
\end{equation}
 $\tilde C(k)$ assumes the smallest value $\tilde C(k_b)$ for given $T^*$ and $\mu^*$ for $k=k_b$ corresponding to
 the minimum of $\tilde V(k)$. 
\begin{figure}[th]
\includegraphics[scale=1]{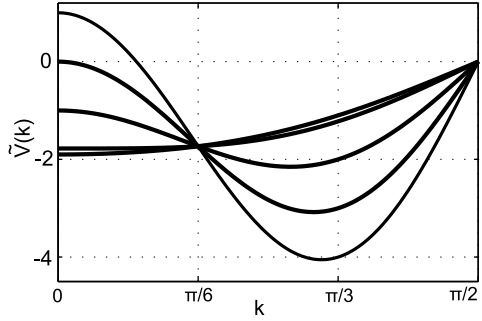}
\caption{$\tilde V(k)$ given by (\ref{V(k)}).  $J^*=0.05,1/9,0.5,1,1.5$ from the bottom to the top line on the left. } 
\label{V_fourier}
\end{figure}
We obtain 
\begin{displaymath}
k_b = \left\{ \begin{array}{ll}
0 & \textrm{ if $J^*<1/9,$}\\
\arccos\sqrt{\frac{1+3J^*}{12J^*}} & \textrm{ otherwise,}\\
\end{array} \right.
\end{displaymath}
and
\begin{displaymath}
\tilde{V}(k_b) = \left\{ \begin{array}{ll}
2(J^{*}-1) & \textrm{ if $J^*<1/9,$}\\
-2J^{*} \Big( \frac{1+3J^{*}}{3J^{*}} \Big)^{3/2} & \textrm{ otherwise.}\\
\end{array} \right.
\end{displaymath}
Let us make a digression on the information on possible types of ordering that is included already in the form of
 $\tilde{V}(k)$. For $k=0$ 
a macroscopic gas-liquid separation occurs (infinite wavelength $2\pi/0$).
For $k>0$ regions of the size $\pi/k$ with excess number of particles,
 separated by depleted  regions of the same size are formed. 
From the form of $\Omega$ (see the first term in Eq.(\ref{wzomeg})) 
it follows 
that  when in the disordered phase  a density wave with the wavelength $2\pi/k$ and the amplitude $|\tilde\phi(k)|$ is
 excited, then the  energy increases by $|\tilde\phi(k)|^2 \tilde{V}(k)/4\pi$.  Note that for $J^*>1$  the energy of the
 homogeneously mixed phase is \textit{lower} than the energy of the system separated into homogeneous gas and liquid phases,
 since $\tilde V(0)=2(J^{*}-1)>0$. Such separation is unfavourable energetically. For 
$1/9<J^*<1$ we have $\tilde V(0)<0$, but  $\tilde{V}(k)$  for $k=0$ assumes a maximum  (Fig.\ref{V_fourier}). It means that the gas-liquid 
separation is energetically favored compared to the homogeneously mixed sample, but the  energy in the presence of
 the density wave with the wavelength $2\pi/k_b$ is lower than in the case of macroscopic separation. 
Competition between the energy and entropy determines whether the periodic ordering or gas-liquid 
separation should occur in this case ($1/9<J^*<1$) for given chemical potential and temperature. 

\begin{figure}[th]
\includegraphics[scale=1]{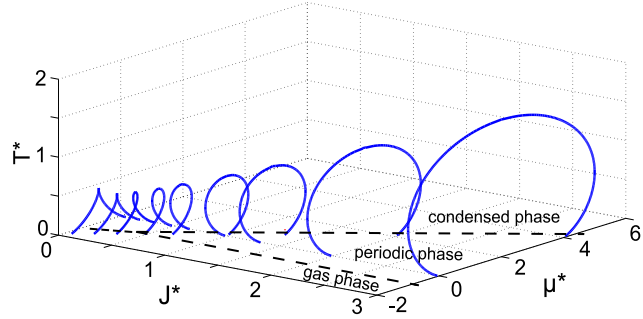}
\caption{Lines of instability (solid) of the homogeneous phase in the $(\mu^*,T^*)$ variables for a range of $J^*$. Similar behavior was obtained in Ref.\cite{barbosa:93:0}.
The coexistence lines at $T^*=0$ are shown as dashed lines.
}
\label{fig_spin}
\end{figure}

The boundary of stability of the disordered phase obtained from $\tilde C(k_b)=0$,
\begin{eqnarray}
 k_{B}T^*=-\tilde V(k_b)\bar\rho_0(1-\bar\rho_0),
\label{rodzina}
\end{eqnarray}
represents the spinodal line of the gas-liquid separation when $k_b=0$ ($J^*<1/9$), and the $\lambda$-line marking the MF 
instability with respect to periodic ordering with the wavelength $2\pi/k_b$ for $k_b>0$ ($J^*>1/9$). It is also interesting to find the $\lambda$-line and the spinodal
 in the $(\mu^*,T^*)$ phase space. From the form of the chemical potential for $\rho=const$, 
$\mu^*=2(J^{*}-1)\rho+T^*\ln[\rho/(1-\rho)]$, we obtain the boundary of stability of the homogeneous phase 
\begin{eqnarray}
\mu^* = (J^{*}-1)(1 \pm q) + T^* \ln \Bigg( \frac{J^{*}}{2T^*}
\Big(\frac{1+3J^{*}}{3J^{*}}\Big)^{3/2} (1 \pm q)^2 \Bigg)
 \end{eqnarray}
where
\begin{eqnarray}
q=\sqrt{1+2 \frac{T^*}{J^{*}}\Big(\frac{3J^{*}}{1+3J^{*}})\Big)^{3/2}}. \quad
\quad
\label{spineq}
 \end{eqnarray}
The shapes of the spinodal and $\lambda$-lines for various $J^*$ are shown in Fig.\ref{fig_spin}. 

In the case of attraction dominated system ($J^*<1/9$), the two branches of the spinodal line separate the region where
 $\Omega$ assumes two minima  for two different constant densities (low-$T^*$ side of the lines) from the region with one 
minimum on the high-$T^*$ side of the lines. At the left branch  of the spinodal the minimum assumed for higher density 
disappears, whereas at the right branch the minimum assumed for the lower density disappears. The phase transition can
 occur on the low-$T^*$ side of the spinodal lines. This is usual behavior associated with the gas-liquid separation. 

For   $1/9<J^*<1$ the branches of the line of instability intersect and form a loop
 for high $T^*$.  Inside the loop  the two minima of $\Omega$ corresponding to constant densities disappear (high-density 
minimum disappears to the left from the branch that corresponds to small $\mu$ for $T^*=0$, and the low-density minimum 
disappears to the right of the branch that  corresponds to large $\mu$ for $T^*=0$). Thus, the homogeneous phase is unstable 
for any density inside the loop. 

Note that when the loop is present, then for decreasing $T^*$ or increasing $\mu^*$ there may exist a sequence of phases:
 disordered - periodic - disordered. Such a sequence agrees with observations of the reentrant melting 
\cite{royall:06:0,elmasri:12:0}.
For the repulsion-dominated system ($J^*>1$)  only the loop is present (Fig.\ref{fig_spin}), and coexistence between homogeneous
 phases is not
expected. 

Our stability analysis is incomplete, because we did not study the boundary of stability of the periodic phase. 
\subsection{Tricritical point}

We assume that when the continuous transition to the homogeneous phase with density $\bar\rho_0(\mu^*,T^*)$ is approached, 
the density in the periodic phase has the form $\bar\rho(x)=\bar\rho_0+\Delta\rho(x)$ with
\begin{eqnarray}
 \Delta\rho(x)=\delta\rho+\phi\cos(k_bx).
\end{eqnarray}
 The continuous transition coincides with the $\lambda$-line, and at the $\lambda$-line $\delta\rho=\phi=0$.
 The  difference between the  grand potential in the periodic and the homogeneous phases, $\Delta\Omega=\Delta\omega L$, 
is a function of $\delta\rho$ and $\phi$ and $\Delta\omega $   can be Taylor expanded for $\delta\rho\to 0$ and $\phi\to 0$. 
When $2\pi/k_b$ is not integer, in calculating $\Delta\omega$ (see (\ref{wzomeg})) we choose $n\to \infty$ such that 
$L\approx 2n\pi/k_b$ and
 make the approximation
\begin{eqnarray}
 \frac{1}{L}\sum_{x=1}^L\cos(k_bx)^m\approx \frac{1}{2\pi}\int_0^{2\pi} \cos z^m dz .
\end{eqnarray}

The second derivative of $\Delta\omega$ with respect to $\phi$ vanishes at the continuous transition,
 while the second derivative 
with respect to $\delta\rho$ is positive. From the extremum condition $\partial\Delta\omega/\partial\delta\rho=0$  we obtain 
\begin{eqnarray}
 \delta\rho=-\frac{A_3(\rho_0)}{4(\beta^*\tilde V(0)+A_2(\rho_0))}\phi^2+O(\phi^4)
\end{eqnarray}
and
\begin{eqnarray}
 \beta\Delta\omega=a_2\phi^2+a_4\phi^4+O(\phi^6)
\end{eqnarray}
where
\begin{eqnarray}
 a_2=\frac{\beta^*\tilde V(k_b)+A_2(\rho_0)}{4}
\end{eqnarray}
\begin{eqnarray}
a_4=\frac{1}{32}\Bigg(
\frac{A_4(\rho_0)}{2}-\frac{A_3(\rho_0)^2}{\beta^*\tilde V(0)+A_2(\rho_0)}
\Bigg)
\end{eqnarray}
and  $A_n(\rho)=d^n\beta f_h(\rho)/d\rho^n$ with $f_h$  given in Eq.(\ref{fh}). 
The transition is continuous for $a_4>0$, and   becomes first order at the tricritical point (TCP) given by $a_2=a_4=0$. 
 We obtain for the density and temperature at the TCP the following expressions
\begin{eqnarray}
 \bar\rho_0^{tcp}=\frac{1}{2}\Bigg[
1\pm\sqrt{\frac{\tilde V(k_b)-\tilde V(0)}{\tilde V(k_b)+3\tilde V(0)}}
\Bigg],
\end{eqnarray}
\begin{eqnarray}
T^*_{tcp}=\frac{-\tilde V(k_b)\tilde V(0)}{\tilde V(k_b)+3\tilde V(0)}.
\end{eqnarray}
Real positive solutions for $\bar\rho_0^{tcp}$ exist for  
$1/9<J^*<1$, ie. when the lines of instability intersect and form a loop. The $T^*_{tcp}$ increases from $0$ for $J^*$ 
decreasing from $1$, and the two TCPs merge at  $\rho=1/2$ and $T^*=4/9$ for $J^*=1/9$,
 i.e. when the disordered fluid starts to be unstable with respect to gas-liquid separation rather than with respect
 to periodic ordering. Thus, the two TCPs merged together change into the standard critical point for $J^*<1/9$.

\subsection{MF phase diagrams}

For very low temperatures the phase diagram is similar to the ground state (Fig.\ref{ground_state}). The gas and liquid phases 
are stable for very low and very high values of $\mu^*$ respectively. If $J^*<1/3$, the two fluid phases coexist. If $J^*>1/3$, the gas coexists with the periodic phase for some value of the chemical potential, and
the periodic phase coexists with the liquid for some larger value of $\mu^*$. The three phases coexist at a triple point 
(TP) for $J^*=1/3$ at $T^*=0$.

Properties of the high $T^*$ part of the phase diagram can be deduced  from the stability analysis. 
When $J^*<1/9$ only standard gas-liquid coexistence is expected. For $J^*>1/9$ there is a continuous transition between 
the fluid and the periodic phase when $T^*>T^*_{tcp}$ and the transition becomes first order when $T^*<T^*_{tcp}$.
 Since for high $T^*$ the periodic phase is stable for $J^*>1/9$ and for $T^*=0$ it is absent for $J^*<1/3$, temperature at the TP increases from $T^*_{tp}=0$ to $T^*_{tp}=4/9$ for $J^*$ decreasing from $1/3$ to $1/9$. Note that for $J^*=1/9$ the TP and the  two TCPs merge, and this special point transforms to the standard critical point for $J^*<1/9$.

For $J^*>1/3$  the TP and the coexistence of the homogeneous phases are absent.  $T^*_{tcp}$ decreases for increasing $J^*$, $T^*_{tcp}=0$ for $J^*=1$ and
the two TCPs  disappear for $J^*>1$. Thus, in the repulsion-dominated systems the high-$T^*$  transition to the periodic phase  is only continuous. This seems to be
inconsistent with the presence of the first-order transition between the periodic and the fluid phases at very low $T^*$.
 This apparent inconsistency follows from the presence of two periodic phases for $J^*>1$. One of them is the same as the phase stable at $T^*=0$. It has large amplitude of density oscillations and the period $l=6$. 
The other phase appears inside the loop of the  $\lambda$-line, has a  period $2\pi/k_b$  and 
 small amplitude of density oscillations. The two periodic phases coexist along the line which is a continuation of the
 coexistence line between the large-amplitude periodic phase and the homogeneous fluid, above the temperature at which this 
transition and the low-$T^*$ branch of the $\lambda$-line (see Fig.3) intersect.
    The coexistence between the two periodic phases terminates at a critical point, where the densities, amplitudes and periods of
 the two phases become the same.  

\begin{figure}[th]
\includegraphics[scale=1]{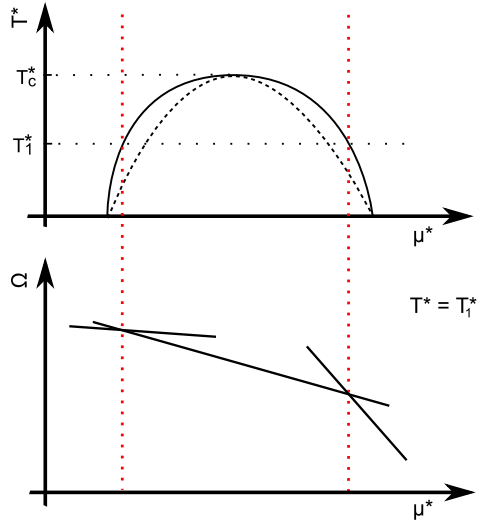}
\caption{Illustration of the method used for obtaining the phase coexistence. In the bottom panel the left, 
central and  right line 
 corresponds to the gas, periodic and disordered liquid phases respectively.
}
\label{figmethod}
\end{figure}

The precise locations of the first-order transition lines have been obtained by calculating the grand potential 
(\ref{wzomeg}) for the average densities that are self-consistent solutions of Eq.(\ref{ror}). 
The method of determining the transition lines is shown  schematically in Fig.\ref{figmethod}. 
We have chosen   three characteristic values of $J^*$, $J^*=3,1/3,1/4$.  $J^*=3$
corresponds to the repulsion dominated system; no TP is present in this case.
For $J^*=1/3$  the TP is present at $T^*=0$, and the repulsion-dominated system crosses over to the attraction-dominated 
system. For  $J^*=1/4$ the TP is present at  $T^*>0$, the two homogeneous phases coexist for low $T^*$, 
and the periodic ordering occurs only at high temperatures.

\begin{figure}[th]
\includegraphics[scale=1]{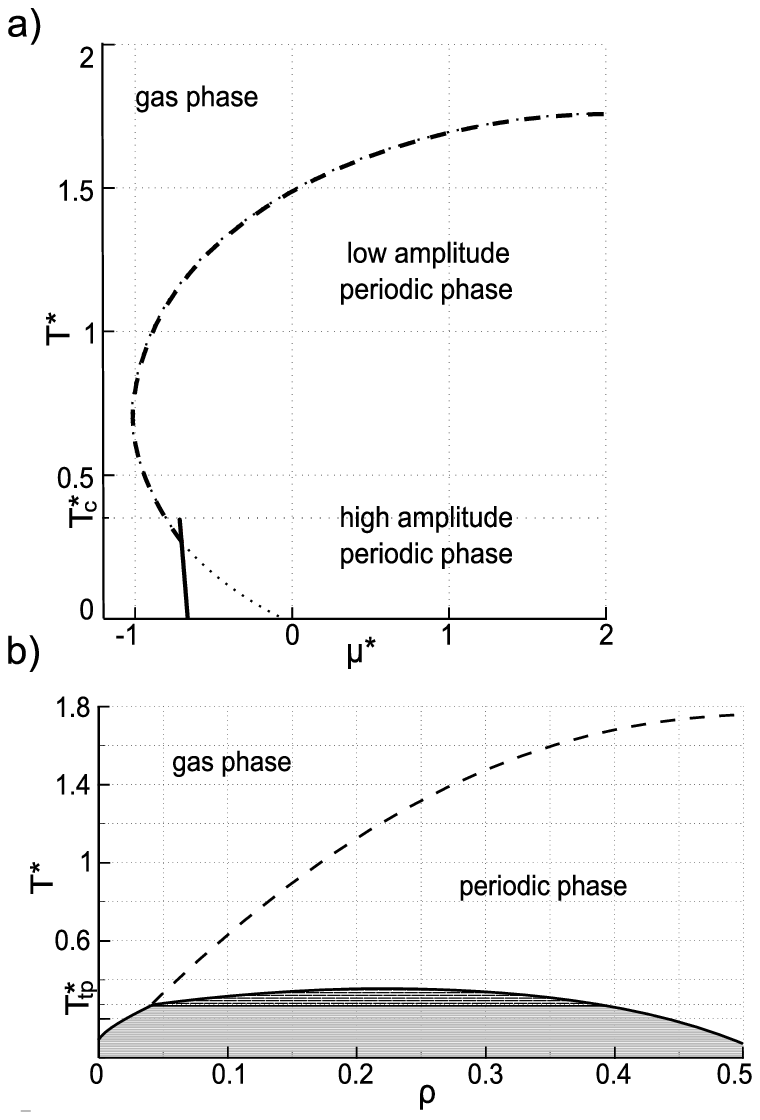}
\caption{ MF phase diagram for $J^* = 3$ in variables $(\mu^*,T^*)$ (a) and $(\rho,T^*)$ (b). The symmetry axis  
is $\mu^*=2$ and $\rho=1/2$ in (a) and (b) respectively.  Only half of the phase diagram is shown because of the symmetry. Dash and solid lines represent continuous  and first-order transitions. The dotted line is the $\lambda$-line.  The two-phase regions 
in (b) are shaded with different shades for different phase equilibria.
The high-amplitude periodic phase coexists with  gas (for $\mu^*<2$ or $\rho<1/2$)  or liquid  (for $\mu^*>2$ or $\rho>1/2$)
   for $T^*<T^*_{tp}$ 
and with the  low-amplitude periodic phase 
for $T^*>T^*_{tp}$.  The line of continuous transitions from the homogeneous to the low-amplitude periodic phase 
terminates at the two first-order transition lines (one for  $\mu^*<2$  or $\rho<1/2$, the other one for $\mu^*>2$ 
 or $\rho>1/2$) at $T^*_{tp}$.
 The coexistence line between the two periodic phases terminates at the critical point with 
$T^*_c \approx 0.34713 $.
}
\label{figMF3}
 \end{figure}

\begin{figure}[th]
\includegraphics[scale=1]{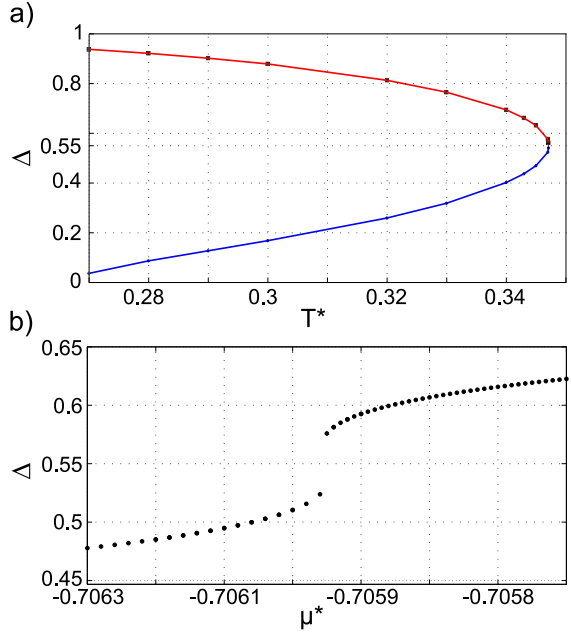}
\caption{Amplitudes of the density profiles in the two periodic phases for $J^* = 3$.  (a)  as a function of temperature along the coexistence line (the lines meet at $T^*_c \approx  0.34713$) (b) as a function of $\mu^*$ for $T^*_c = 0.347$.
}
\label{figMFampl}
 \end{figure}

\begin{figure}[th]
\includegraphics[scale=1]{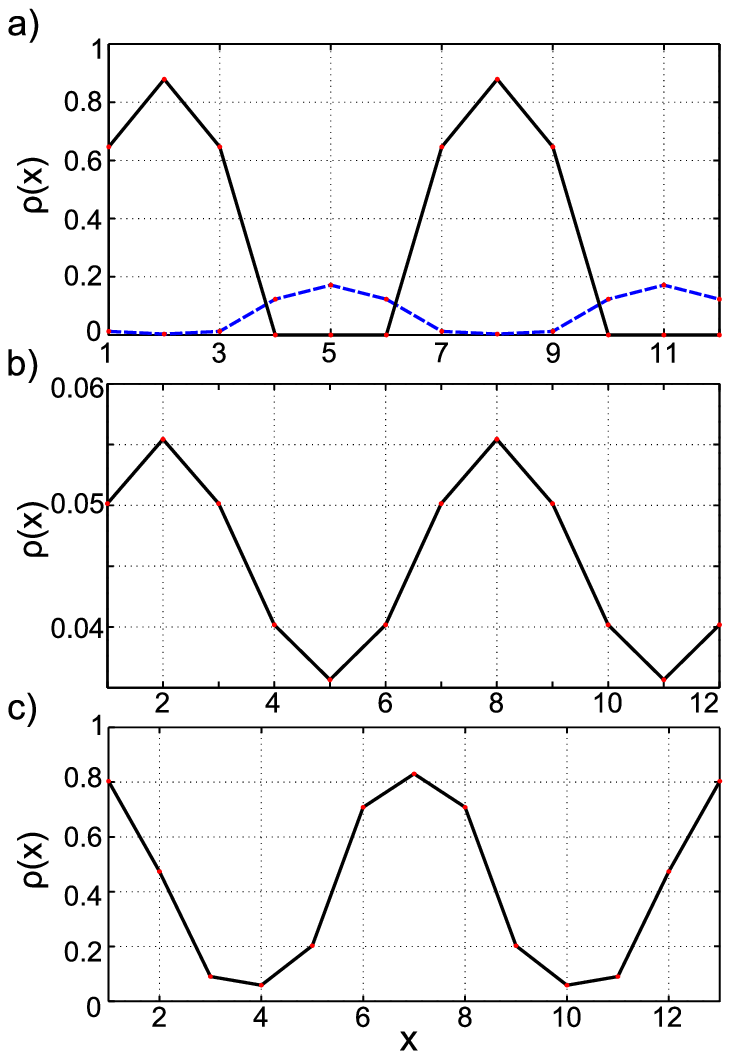}
\caption{ Density profiles  (a) in the coexisting  high- and low-amplitude phases for $T^*=0.3$. 
The lines are shifted horizontally for clarity.  (b) in the low-amplitude phase
 close to the continuous transition to the fluid at  $T^*=0.3$ and (c) for
 $T^*=1.3$ and $\mu^*=1$. The  quasi-periodic structure with
 a period incommensurate with the lattice is obtained from a density profile with a large-period when $2\pi/k_b$ is noninteger.
The lines connecting the results for integer $x$ are to guide the eye.
}
\label{figMFdp}
 \end{figure}

The $(\mu^*,T^*)$ and $(\rho,T^*)$ phase diagrams  for $J^*=3$ are shown in Fig.\ref{figMF3}. 
 The amplitudes of the two
periodic phases along their coexistence line  and for $T^*=0.347$ are shown in Fig.\ref{figMFampl} 
as functions of $T^*$ and $\mu^*$. The density profiles in the two periodic phases for selected thermodynamic 
states are shown in Fig.\ref{figMFdp}. As far as we know, coexistence of two ordered phases with the same symmetry but 
diffrent degree of order has not been reported yet. 

 
\begin{figure}[th]
\includegraphics[scale=1]{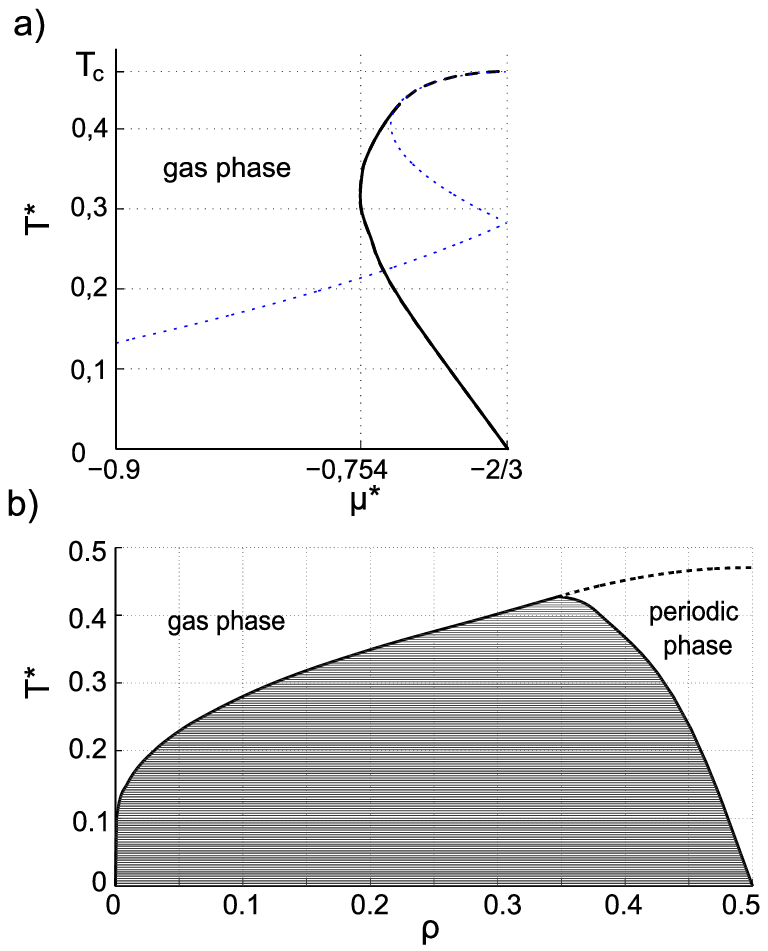}
\caption{ MF phase diagram for $J^* = 1/3$ in variables $(\mu^*,T^*)$ (a) and $(\rho,T^*)$ (b). The symmetry axis
 is $\mu^*=-2/3$ and $\rho=1/2$ in (a) and  (b) respectively. Only half of the phase diagram is shown because of the symmetry.
 Dash and solid lines represent continuous
 and first-order transitions between the disordered fluid and the periodic phase. The dotted line is
the $\lambda$-line. The periodic phase in (a) is stable inside the loop (thick line). 
The two-phase regions in (b) are shaded. }
\label{figMF1/3}
 \end{figure}

The $(\mu^*,T^*)$ and $(\rho,T^*)$ phase diagrams $J^*=1/3$ are shown in Fig. \ref{figMF1/3}.  
In this case  there is a single periodic
 phase that coexists with the dilute and the dense homogeneous fluid for $T^*<T^*_{tcp}$ and undergoes a continuous
 transition to the fluid for $T^*>T^*_{tcp}$. Similar phase behavior was obtained in Ref.\cite{archer:08:1}, 
where only one-dimensional density oscillations were assumed in the Landau-type and density-functional theories.
 
\begin{figure}[th]
\includegraphics[scale=1]{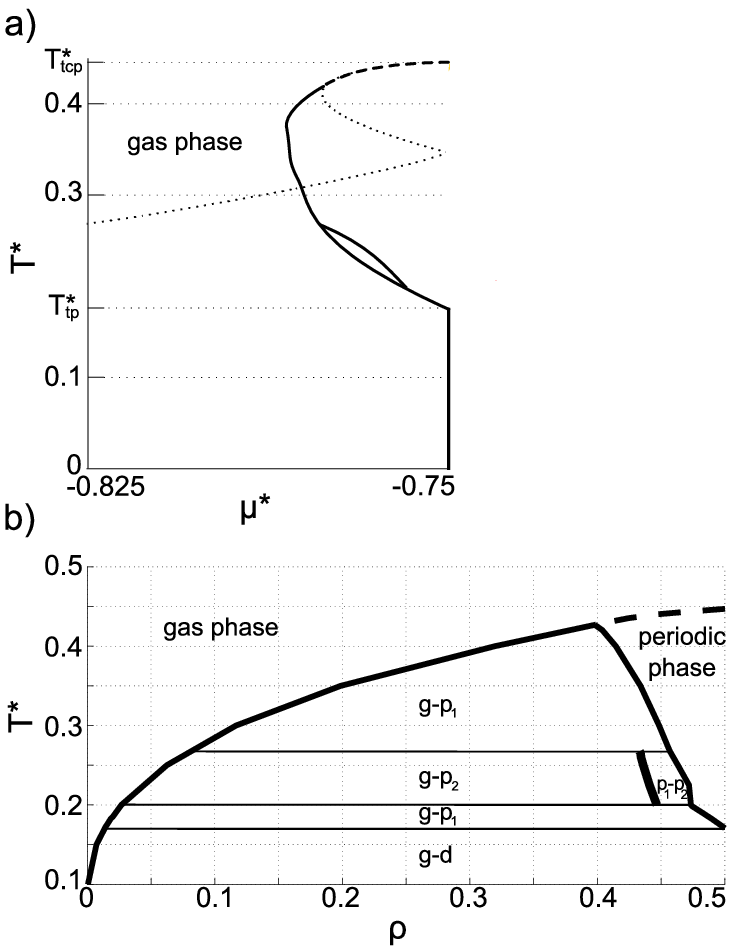}
\caption{ MF phase diagram for $J^* = 1/4$ in variables $(\mu^*,T^*)$ (a) and $(\rho,T^*)$ (b). 
The symmetry axis in (a) is $\mu^*=-3/4$ and $\rho = 1/2$ in (b).  Only half of the phase diagram is shown because of the symmetry. Dash and solid lines represent continuous
 and first-order transitions. The dotted line is the $\lambda$-line. The coexisting phases in the two-phase regions in (b) are labeled by $g$ for gas, $d$ for dense fluid and $p_1$, $p_2$ 
for the periodic phases with the smaller and the larger period respectively. The density range of stability of the 
large-period phase is within the thickness of the line. The periodic phases in (a) are stable inside the lense ($p_2$) and inside the loop ($p_1$).  }
\label{figMF1/4}
 \end{figure}

The $(\mu^*,T^*)$ and $(\rho,T^*)$ phase diagrams for $J^*=1/4$ are shown in Fig. \ref{figMF1/4}. 
In this case the two fluid phases
 coexist at $T^*<T^*_{tp}\approx 0.17063$, while for $T^*>T^*_{tp}$ each of them coexists with the periodic phase up 
to $T^*_{tcp}$
 where the transition becomes continuous. For some range of temperature another periodic phase, with the same amplitude but
 with a larger period (hence lower density)  becomes stable. This phase diagram is similar to the phase diagram obtained in Ref.\cite{archer:08:1} for very weak repulsion. Moreover,  when electrostatic repulsion is added to the Landau functional, similar phase diagram is obtained \cite{andelman:87:0}.
\section{Exact solutions}
\subsection{Transfer matrix and exact expressions}
Since the range of interactions is $3$, we coarse-grain the lattice and introduce $L/3$  boxes consisting of three neighboring 
lattice sites, and labeled by integer $1\le k\le L/3$.
The microstates in the $k$-th box are  
 \begin{eqnarray}
\label{S(k)}
\hat S(k)=(\hat\rho(3k-2),\hat\rho(3k-1),\hat\rho(3k)). 
 \end{eqnarray}
There are $2^3$ possible microstates in each box. 
We distinguish 4 states with the first site occupied and the remaining sites either occupied or empty,
 and denote such states by $\hat S_1(k)=(1,\hat\rho(3k-1),\hat\rho(3k))$. Likewise, we denote states with the second 
  site  occupied by $\hat S_2(k)=(\hat\rho(3k-2),1,\hat\rho(3k))$,  and with the third  site  occupied by
 $\hat S_3(k)=(\hat\rho(3k-2),\hat\rho(3k-1),1)$. 

The Hamiltonian of the system with PBC can be written in the form
 \begin{eqnarray}
  H^*[\{\hat\rho(x)\}]=\sum_{k=1}^{L/3} H^*_t[\hat S(k),\hat S(k+1)]
 \end{eqnarray}
where  
\begin{eqnarray}
 H^*_t[\hat S(k),\hat S(k+1)]=\sum_{x=3k-2}^{3k}\big[
-\hat\rho(x)\hat\rho(x+1)+J^*\hat\rho(x)\hat\rho(x+3)-\mu^*\hat\rho(x)
\big].
\end{eqnarray}
We introduce a $8\times 8$  transfer matrix ${\bf T}$ with the matrix elements  
\begin{eqnarray}
\label{T}
 T(\hat S(k),\hat S(k+1))\equiv e^{-\beta^* H^*_t[\hat S(k),\hat S(k+1)]}=
\sum_{i=1}^8P_i(\hat S(k))\lambda_iP_i^{-1}(\hat S(k+1)),
\end{eqnarray}
where the eigenvalues of ${\mathbf T}$ are denoted by $\lambda_i$ such that
$|\lambda_i|\ge|\lambda_{i+1}|$, 
the elements $(\hat S,i)$ of the  matrix ${\mathbf P}$ transforming  ${\bf
T}$ to its eigenbasis  are denoted by $P_i(\hat S)$,
 and 
the elements  $(i,\hat S)$ of the matrix inverse to ${\bold P}$ by $P_i^{-1}(\hat S)$.
Note that ${\bf T}$ is not symmetric, hence  pairs of complex-conjugate eigenvalues may occur.  
 However, because ${\bf T}$
is a finite matrix with positive elements, from the Frobenius  theorem it follows that the largest  (in absolute value) 
eigenvalue is  non-degenerate. 

The grand statistical sum in terms of the transfer matrix takes the form
\begin{eqnarray}
\label{Xi}
 \Xi=Tr {\bf T}^{L/3} =\sum_{i=1}^8\lambda_i^{L/3}
\end{eqnarray}
and for the grand potential we obtain
\begin{eqnarray}
\label{p}
 \Omega^*/L=-p^*=-T^*\Bigg[\frac{1}{3}\ln \lambda_1+\frac{1}{L}\ln\Big(
1+\sum_{i=2}^8\Big(\frac{\lambda_i}{\lambda_1}\Big)^{L/3}\Big)
\Bigg]\simeq_{L\to\infty} -\frac{T^*}{3}\ln \lambda_1.
\end{eqnarray}
 In the case of PBC the average density is independent of the position, $\bar\rho=\langle\hat\rho(1)\rangle$.
From  the definition of the average density and from  Eqs. (\ref{T}) and (\ref{Xi}) we obtain
\begin{eqnarray}
\label{<rho>}
 \langle\hat\rho(1)\rangle=\frac{\sum_{\hat S_1(1)} \sum_{i=1}^8\lambda_i^{L/3}P_i^{-1}(\hat S_1(1))P_i(\hat S_1(1))}
{\sum_{i=1}^8\lambda_i^{L/3}}\simeq_{L\to\infty}\sum_{\hat S_1(1)} P_1^{-1}(\hat S_1(1))P_1(\hat S_1(1)).
\end{eqnarray}

 Let us consider the correlation function
for the sites separated by a distance $x=3k+i$, where $ k\ge 0$ and $i=0,1,2$. 
Because the interaction range is $3$, and the transfer matrix operates between triples of sites, we shall obtain
different expression for $\langle \hat\rho(1)\hat\rho(1+x)\rangle$ for different $i=0,1,2$. We introduce the notation
\begin{eqnarray}
 G(3k+i)=\langle \hat\rho(1)\hat\rho(1+3k+i)\rangle -\langle \hat\rho(1)\rangle^2.
\end{eqnarray}
From the definition of $\langle \hat\rho(x)\hat\rho(x')\rangle$ and from Eqs. (\ref{T}) and (\ref{<rho>}) 
we obtain  the  asymptotic expression for $L\to\infty$
\begin{eqnarray}
 G(3k+i)=\sum_{n=2}^8\Big(
\frac{\lambda_n}{\lambda_1}
\Big)^{k}A_1^{(n)}B_{1+i}^{(n)}
\end{eqnarray}
where $i=0,1,2$,
\begin{eqnarray}
 A_j^{(n)}=\sum_{\hat S_j}P_n(\hat S_j)P_1^{-1}(\hat S_j),
\end{eqnarray}
\begin{eqnarray}
 B_j^{(n)}=\sum_{\hat S_j}P_n^{-1}(\hat S_j)P_1(\hat S_j),
\end{eqnarray}
and $\hat S_j$ is defined below Eq.(\ref{S(k)}).
The asymptotic decay of correlations for $k\gg 1$ is determined by the eigenvalue $\lambda_2$ 
 with the second largest absolute value. 

If  $\lambda_2$ is real, then  for $k\gg 1$ and $i=0,1,2$ we can write
\begin{eqnarray}
 G(3k+i)=(sgn(\lambda_2))^k e^{-3k/\xi} A_1^{(2)}B_{1+i}^{(2)}
\end{eqnarray}
where the correlation length is
\begin{eqnarray}
\label{xi}
\xi=3/\ln \Big(\frac{\lambda_1}{|\lambda_2|}
\Big).
\end{eqnarray}
Note the qualitatively different behavior for $\lambda_2>0$ and $\lambda_2<0$. For $\lambda_2<0$  the 
correlation function changes sign when  the separation between the particles increases by 3, in analogy with the density of
 the periodic phase in the ground state. The case $\lambda_2>0$ corresponds to decay of correlations in the gas 
or liquid phases where no clusters consisting of three particles separated by 3 vacancies are formed.

If $\lambda_2$ is complex, then $\lambda_3=\lambda_2^{*}$, $A_1^{(3)}=A_1^{(2)*}$ and  $B_{j}^{(3)}= B_{j}^{(2)*}$. We 
  introduce the notation
\begin{eqnarray}
 \lambda_2=|\lambda_2|e^{i\lambda}, \hskip1cm A_1^{(2)}=|A_1^{(2)}|e^{i\alpha_1},\hskip1cmB_{j}^{(2)} =|B_{j}^{(2)} |e^{i\beta_j},
\end{eqnarray}
and  for $k\gg 1$ and $i=0,1,2$ obtain the asymptotic expression
\begin{eqnarray}
\label{corr1}
  G(3k+i)= {\cal A}_i e^{-3k/\xi}\cos\big(
k\lambda +\theta_i
\big)
\end{eqnarray}
where  ${\cal A}_i=2|A_1^{(2)}||B_{1+i}^{(2)}|$ and
$\theta_i=\alpha_1+\beta_{1+i}$, $i=0,1,2$.
 Similar expression was proposed in Ref.\cite{pini:06:0} for  a 3d system. The structure factor obtained
in experiments and theory \cite{stradner:04:0,shukla:08:0,sear:99:0,pini:00:0,imperio:04:0,imperio:07:0,imperio:06:0,pini:06:0,archer:07:0,archer:07:1,ciach:08:1,archer:08:0,archer:08:1} 
is also consistent with this form.
In general, $-\pi\le\lambda\le \pi$, and $2\pi/\lambda$  is noninteger. Except from $\lambda=\pm \pi$ (but in this case the imaginary part of $\lambda_2$ vanishes), the period of the exponentially 
damped oscillations is incommensurate with the lattice. This is similar to the results of the MF stability analysis and to the 
 incommensurate density profiles obtained in MF for  higher temperatures.

When  $\lambda=\pm \pi\mp\epsilon$ with $\epsilon  \ll 1$, then we can write Eq.(\ref{corr1}) in the equivalent form
\begin{eqnarray}
\label{Ge}
  G(3k+i)=(-1)^{k}e^{-3k/\xi}{\cal G}(k,i)
\end{eqnarray}
with 
\begin{eqnarray}
\label{Gea}
  {\cal G}(k,i)={\cal A}_i\cos(6\pi k/w+\phi_i),
\end{eqnarray}
where the phase and the period of the amplitude modulations are $\phi_i=-\theta_isgn(\lambda)$ and
\begin{eqnarray}
\label{w}
w=\frac{6\pi}{|\lambda-sgn(\lambda)\pi|}.
\end{eqnarray}
The first factor in (\ref{Ge})  changes sign when the distance increases by $3$. The last factor describes the 
modulated amplitude  with the wavelength of modulations $w\gg 6$ if $\lambda\to \pm\pi$.

We have obtained $\lambda_i$ and the matrix ${\bold P}$ numerically for different $J^*,\mu^*$ and $T^*$  and the results are presented  in the next section.

\subsection{Results}
There are no phase transitions in a thermodynamic sense in one-dimensional systems. However, instead of a discontinuity,
 a rapid change
of the density as a function of $\mu^*$ or $p^*$ can occur.  Moreover, instead of long-range order and the associated
 periodic density, a short-range order with exponentially damped
 oscillatory decay of correlations with very large correlation length may exist. In order to verify if  such pseudo 
phase transitions occur in this model,  we  calculate density and pressure for several values of $J^*$ for the
 range of $\mu^*$ and $T^*$ corresponding to  the phase transitions obtained in MF.  In the next step we  examine 
the correlation functions.

\subsubsection{Thermodynamic properties (Equation of state)}

\begin{figure}[th]
\includegraphics[scale=1]{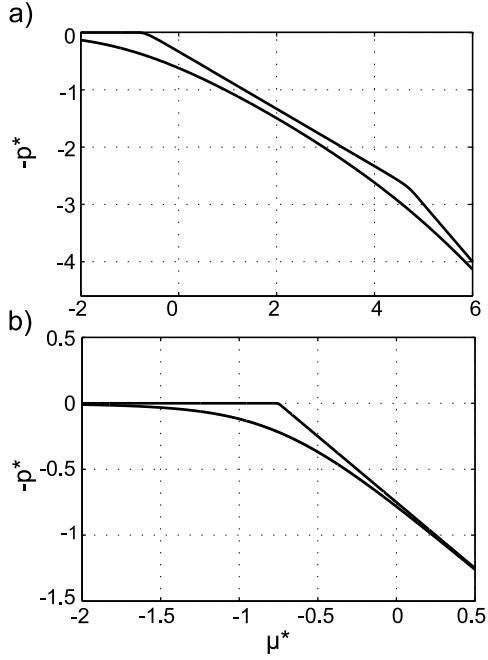}
\caption{$p(\mu^*)$ obtained from Eq.(\ref{p}) (a) $J^*=3$; top line:  $T^* = 0.1$, bottom line: $T^*= 1$  and  (b) $J^*=1/4$; top line: $T^*=0.05$, bottom line: $T^*=0.5$.
}
\label{fig_p(mu)}
\end{figure}

\begin{figure}[th]
\includegraphics[scale=1]{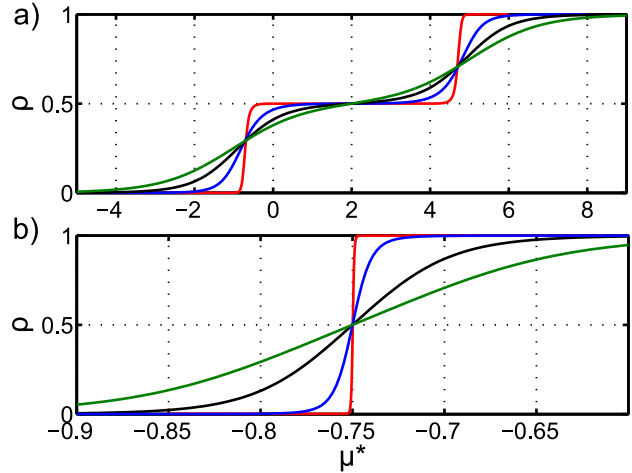}
\caption{$\rho(\mu^*)$ obtained from Eq.(\ref{<rho>}) for $J^*=3$ and $T^*=0.1,0.4,0.7,1$ (top to bottom line on the right) (a) and $J^*=1/4$ and $T^*=0.005,0.05,0.1,0.15$ (top to bottom line on the right) (b).
}
\label{fig_rho(mu)}
\end{figure}

\begin{figure}[th]
\includegraphics[scale=1]{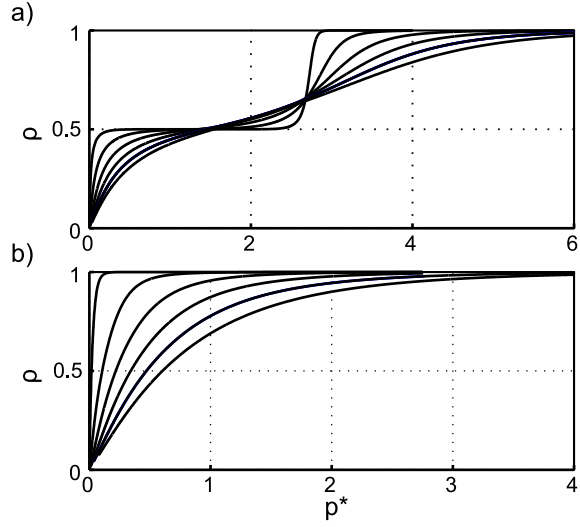}
\caption{EOS $\rho(p^*)$ isotherms obtained from Eqs.(\ref{p}) and (\ref{<rho>}) for $T^*=0.1,0.2,0.3,0.4,0.5$ and $1$ (top to bottom line on the left) for $J^*=3$  (a) and $J^*=1/4$ (b).
}
\label{figEOS}
\end{figure}

\begin{figure}[th]
\includegraphics[scale=1]{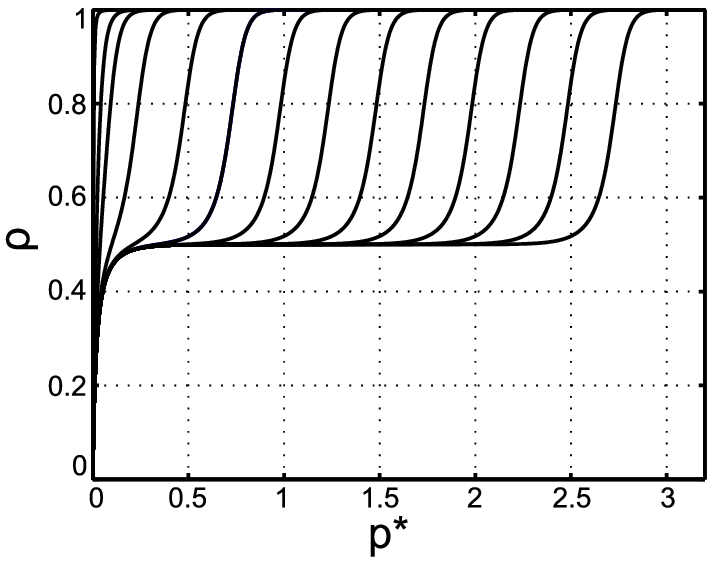}
\caption{EOS $\rho(p^*)$ isotherms obtained from Eqs.(\ref{p}) and (\ref{<rho>}) for $T^*=0.1$. 
From the left to the right line $J^*=0.1, 0.25, 1/3, 0.5, 0.75,...,2.75,3$.
}
\label{figEOS1}
\end{figure}

 In Figs. \ref{fig_p(mu)}-\ref {fig_rho(mu)} $p(\mu^*)$ and $\rho(\mu^*)$ obtained from Eqs.(\ref{p}) and (\ref{<rho>})
 are shown for $J^*=3$ and $J^*= 1/4$ 
for $0.05<T^*<1$.  By eliminating $\mu^*$ from Eqs. (\ref{p}) and (\ref{<rho>}) we obtain the EOS, and present several
 isotherms $\rho(p^*)$ in Figs.\ref {figEOS} and \ref {figEOS1}. The chosen strengths of the repulsion to attraction ratio
 correspond to qualitatively 
different ground state and MF phase diagrams 
(see Figs.1, \ref{figMF3} and \ref{figMF1/4}). 
Let us first discuss $J^*=3$. 
For low $T^*$ one can observe that although $p^*(\mu^*)$ is a smooth function, its slope changes rapidly for the two values
 of $\mu^*$ that correspond to the phase transitions at $T^*=0$ and in MF. 
In accordance with this behavior the density changes from nearly $0$ 
to $1/2$ in a very narrow range of $p^*$ and $\mu^*$, remains nearly constant for large intervals of  $p^*$ and $\mu^*$, and 
again 
 changes rapidly from $\rho=1/2$ to $\rho\approx 1$. Very large compressibility for $\rho\ne 1/2$ changes to very small 
compressibility for $\rho\approx 1/2$.
 It is necessary to substantially  increase the pressure 
in order to induce a slight increase of the density from  $\rho=1/2$, and a slight further increase of pressure 
is sufficient for a rapid compression  to $\rho\approx 1$. 
 When $T^*$ increases from $T^*\approx 0.1$, the density changes from the
 gas density 
to $1/2$ more and more gradually. For $T^*>0.5$ there are no abrupt changes of the  slopes of the $\rho(p^*)$ and $\rho(\mu^*)$
lines,
 but the curvature of
 these lines is significantly smaller than in the one-phase region of a simple fluid. This is because  the repulsion between 
the particles at the  distance $3$ leads to a significant increase of pressure for random distribution of particles.
 On the other hand, 
small pressure for $\rho<1/2$ when  $T^*$ is low signals that in majority of states clusters made of at most 3 particles are
 separated by at least 3 empty sites.
Similar behavior is observed for $1<J^*<3$ (Fig.\ref{figEOS1}), but the range of $\mu^*$,  $p^*$ and $T^*$ for which $\rho\approx 1/2$ and remains nearly constant decreases with decreasing $J^*$. For $J^*<1 $  the plateau at the $\rho(\mu^*)$ and $\rho(p^*)$ curves for $\rho=1/2$  disappears.


For $J^*=1/4$ we can see in Fig. \ref{fig_p(mu)} b) a rapid change of the slope of the $p^*(\mu^*)$ line and in Fig. \ref{fig_rho(mu)}
the corresponding change of density from $\rho\approx 0$ to $\rho\approx 1$ 
when $T^*<0.06$. For  $T^*>0.15$ 
 the shapes of $p^*(\mu^*), \rho(\mu^*)$ and $\rho(p^*)$ (Fig.\ref{figEOS}) resemble the corresponding curves in the single-phase simple fluid.
  We thus see a pseudo-transition 
between the gas and liquid phases for very low $T^*$.  We conclude that the thermodynamic properties
 show no signature  of the weakly-ordered periodic phase previously found in MF (Fig.\ref{figMF1/4}). 
By comparing Figs.\ref{figEOS} a) and b)  one can see
 the much lower pressure in this case than for the repulsion-dominated case of $J^*=3$.

 \subsubsection{Structure (correlation function)}

 Our aim in this section is to discuss the exact results for the correlation function 
for $J^*=3$ and $J^*=1/4$, corresponding to qualitatively different ground state (Fig.\ref{ground_state}) and MF phase diagrams (Figs.\ref{figMF3} and \ref{figMF1/4}). We particularly address the question
 for what parameters the periodic order occurs, and how the range and amplitude of the correlation function depends on $\mu^*$, $T^*$ and $J^*$.

\begin{figure}[th]
\includegraphics[scale=1]{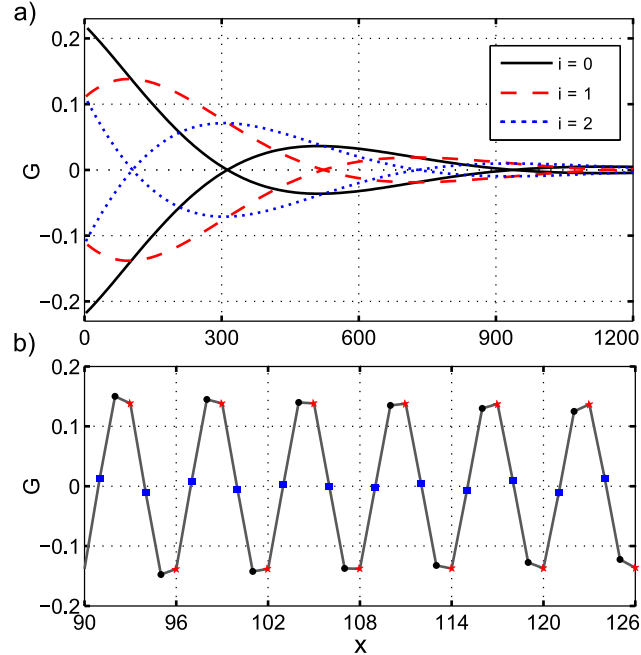}
\caption{The correlation function $G(x)$ for $x=3k+i$ with $i=0,1,2$
(Eq.(\ref{corr1})) for $J^*=3$,  $\mu^*=0$ and $T^*=0.1 $ (inside the MF
stability region of the periodic phase).  Solid line and the circles (black), dashd line and the asterisks
(red) and dotted line and the squares  (blue)  correspond to
$i=0,1,2$ respectively. The bottom panel shows a small portion of the upper
panel.
}
\label{figcor3i}
\end{figure}

\begin{figure}[th]
\includegraphics[scale=1]{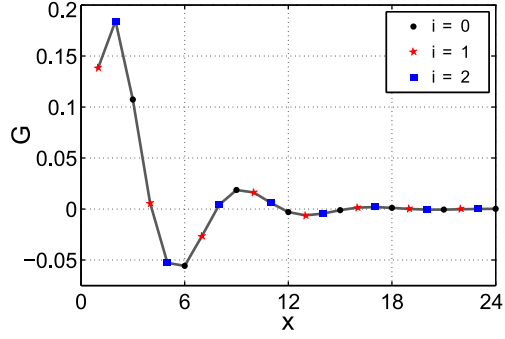}
\caption{The correlation function  $G(x)$ for $x=3k+i$ with $i=0,1,2$ (Eq.(\ref{corr1})) for $J^*=3$,  $\mu^*=-0.7$ and $T^*=0.1 $ (outside the MF stability region of the periodic phase). Black (circle), red (asterisk) and blue (square) symbols correspond to $i=0,1,2$ respectively.
}
\label{figcor3o}
\end{figure}

For $J^*=3$ we obtain complex $\lambda_2$ for the considered region of $(\mu^*,T^*)$. In this case the correlation function is given in Eq.(\ref{corr1}), and presented in Figs.\ref{figcor3i} and \ref{figcor3o}. In Fig.\ref{figcor3i}   $\mu^*$ corresponds to $\rho\approx 1/2$, where the periodic phase is predicted in MF,  and  in Fig.\ref{figcor3o}  $\mu^*$ corresponds to $\rho\approx 0$ (homogeneous gas in MF).
 

\begin{figure}[th]
\includegraphics[scale=1]{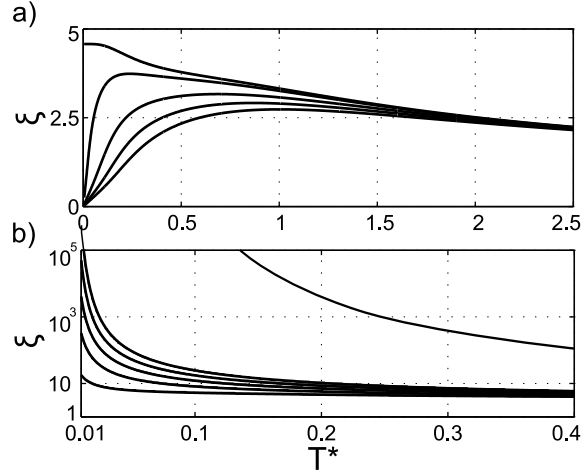}
\caption{The correlation length $\xi$ (Eq.(\ref{xi})) for $J^*=3$ as a function
of  $T^*$ (a) outside the MF stability region of the periodic phase. From the
top to the bottom line $\mu^* = -2/3,-0.7,-0.8,-0.9,-1$  and (b) inside the MF
stability region of the periodic phase.  From the  bottom to the top line $\mu^*
= -0.65, -0.6, -0.55,-0.5,-0.45$ and $2$.
}
\label{figxi}
\end{figure}

\begin{figure}[th]
\includegraphics[scale=1]{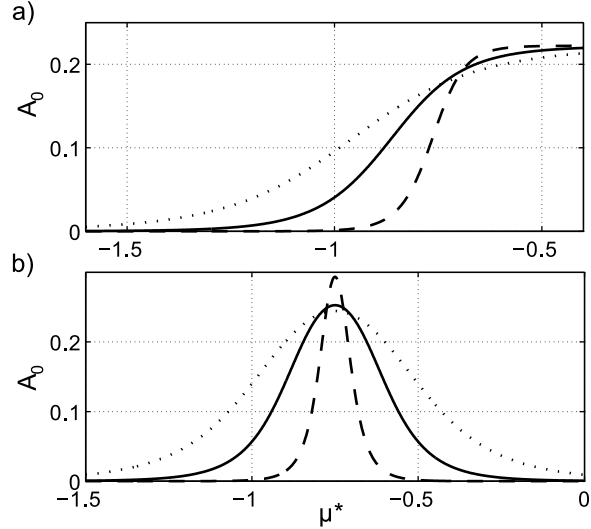}
\caption{The amplitude ${\cal A}_0$ of the correlation function (see Eq.(\ref{corr1}) and below)  as a function of  
$\mu^*$. Dash, solid and dotted lines correspond to  $T^*=0.1,0.2,0.3$  respectively. $J^* = 3$ (a) and $J^*=1/4$ (b).
}
\label{figamp3}
\end{figure}

\begin{figure}[th]
\includegraphics[scale=1]{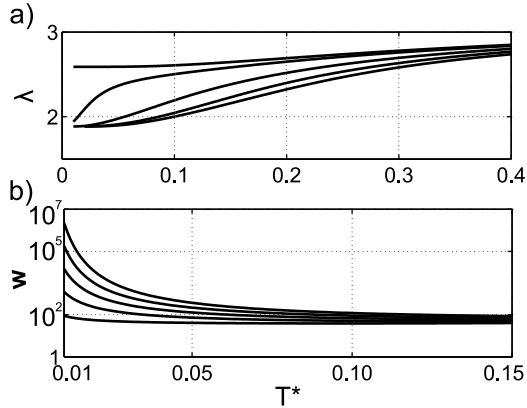}
\caption{(a) the wavenumber $\lambda$ of the correlation function (Eq.(\ref{corr1})).
 $J^*=3$ and  $\mu^* =
-2/3,-0.7,-0.8,-0.9,-1$ from the top to the bottom line  (b) the period $w$ (Eqs.(\ref{Ge})-(\ref{w})) of the
amplitude modulations. $J^*=3$ and $\mu^*=
-0.65, -0.6, -0.55,-0.5,-0.45$ from the bottom to the top line.
}
\label{figw}
\end{figure}

The correlation length $\xi$ (Eq.(\ref{xi})) and the
amplitude of the correlation function ${\cal A}_0$  (see below Eq.(\ref{corr1})) are shown in Figs.\ref{figxi} and \ref{figamp3} respectively.  For $\mu\le -2/3$  the wavenumber $\lambda$ is shown in Fig.\ref{figw} a), and  for $\mu\ge -2/3$ the period  $w$ of modulations of the amplitude (see (\ref{Ge}) and (\ref{Gea})) is shown in Fig.\ref{figw} b).

From Fig.\ref{figxi} a) it follows that for $\mu^*< -2/3$  the correlation length $\xi$ first increases slightly for 
decreasing $T^*$, but starting form $T^*$ depending on $\mu^*$ decreases rapidly to $0$ for $T^*$ decreasing to $0$.
 Analogous  behavior is predicted for $\mu^*> 14/3$ by the model symmetry. For $-2/3<\mu^*<14/3$ (stability of the
 periodic phase for $T^*=0$) the  correlation length increases for decreasing $T^*$. For given $T^*$ the correlation
 length increases with increasing $\mu^*$ when $\mu^*<2$  and assumes a maximum for $\mu^*=2$.  The maximum of $\xi$ 
is  very large for $T^*<0.15 $. For $\xi\sim 10^5$ the range of the 'short-range order` is in fact macroscopic. 
For particles with a diameter $10nm$ the periodic arrangement persists to distances $\sim 1 mm$. 
The amplitude of the correlation function for $T^*\le 0.1$  increases sharply  from a very small value for $\mu^*<-1$ to
 $\sim 0.2$ for $\mu^*>-0.7$ (see Fig.\ref{figamp3}a). The period $w$ of the modulations of the correlation function
(Eqs.(\ref{Ge})- (\ref{w}))  increases for decreasing $T^*$, indicating more and more ordered structure (see Fig. \ref{figw}b). All these results confirm a qualitative change of the structure along the lines $\mu^*=-2/3,14/3$ for low $T^*$. For $-2/3<\mu^*<14/3$ quasi long-range order with the very large correlation length and the amplitude that for low $T^*$ rapidly decreases at the boundaries of this region exists. From Figs.\ref{figxi} and   \ref{figamp3} we can see that the increasing correlation length and
 amplitude for increasing $T^*$ when $-1<\mu^*<-2/3$ indicates a change from a less to a more ordered structure when
 $T^*$ increases.  The rapid increase of the amplitude as a function of $\mu^*$ for $\mu^*\approx-2/3$  near $T^*$ 
corresponding to the maximum of $\xi$  resembles the transition between the  periodic phases with weak and strong order
 found in MF (compare Figs. 17a and  \ref{figMFampl}b).  When $T^*$ further increases, the properties of the correlation 
function change more gradually and the correlation length becomes short, in consistency with the continuous  transition 
between the ordered and disordered phases found in MF  for high $T^*$. 
Despite the absence of the phase transition in the strict sense, we can see a change from a quasi-ordered periodic 
structure to  the structure with much lower degree of order.  
\begin{figure}[th]
\includegraphics[scale=1]{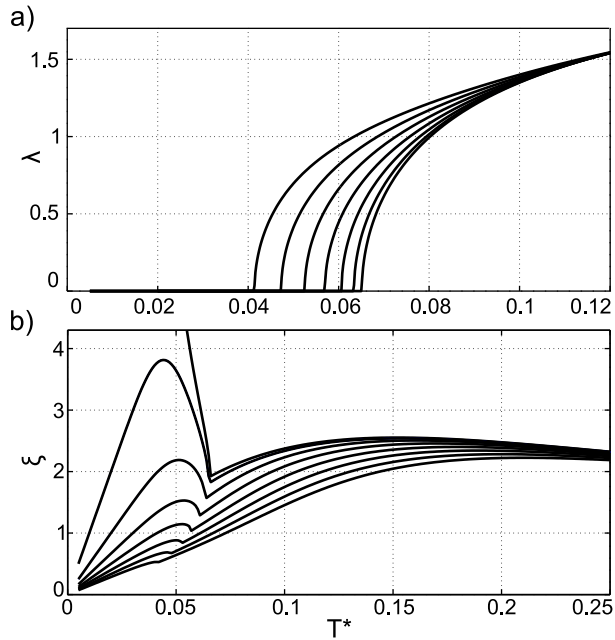}
\caption{The wavenumber $\lambda$  (a)  and the correlation length $\xi$ (Eq. (\ref{xi})) (b) of the correlation function 
 (Eq.(\ref{corr1}))  as a function of $T^*$ for $J^*=1/4$. From the bottom to the top line in (a) and from the top to 
the bottom line in (b)  $\mu^*=-0.75,-0.75\pm 0.01,-0.75\pm 0.02,\ldots,-0.75\pm 0.07$. 
}
\label{figxi1/4}
\end{figure}

\begin{figure}[th]
\includegraphics[scale=1]{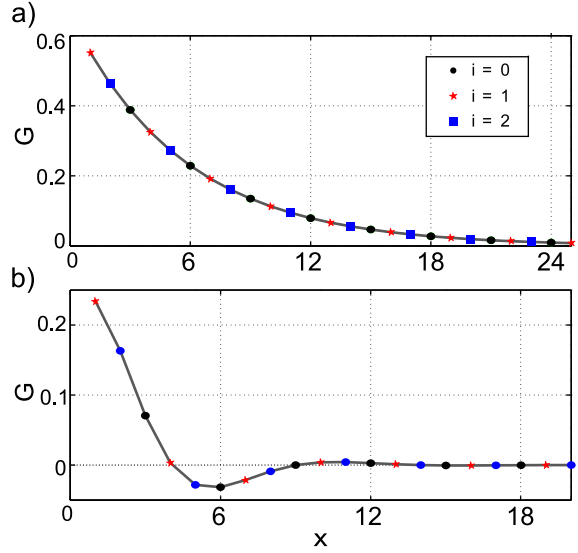}
\caption{The correlation function  $G(x)$ for $x=3k+i$ with $i=0,1,2$ (Eq.(\ref{corr1})) for  $J^*=1/4$ and  
$\mu^*=-3/4$  for $T^*=0.05$ (a) and $T^*=0.2$ (b). Black (circle), red (asterisk) and blue (square) 
symbols correspond to $i=0,1,2$ respectively.
}
\label{figcor1/4}
\end{figure}

For $J^*=1/4$ we obtain $\lambda_2$ that is a real number on the  low-$T^*$ side of a line $T^*_{cross}(\mu^*)$, and a 
complex number for $T^*$ above this line. As a result, a monotonic decay of correlations is found for  
$T^*<T^*_{cross}(\mu^*)$, and an oscillatory decay sets in for $T^*>T^*_{cross}(\mu^*)$. The
derivatives of the  correlation length $\xi$  and the wavenumber $\lambda$ with respect to $T^*$ have a
 discontinuity when the imaginary part of $\lambda_2$ appears (see Fig.\ref{figxi1/4}). 
 The amplitude of the correlation function (Fig.\ref{figamp3}b) increases from a very small value for $\mu^*<-1$ to
 $\sim 0.3$
 for $\mu^*=-0.75$ for $T^*=0.1$. For larger $T^*$ the increase of the amplitude is more gradual. 
The correlation function 
shown in Fig. \ref{figcor1/4} confirms that for low $T^*$ the correlations decay monotonically,
 whereas for higher $T^*$ the oscillatory decay of correlations is present. The correlation length, however,
 is rather short, as shown in Fig.\ref{figxi1/4}. The monotonic decay of correlations for low $T^*$,
 and oscillatory decay of correlations at higher $T^*$ together with a rather rapid increase of the amplitude 
of the correlation function from a very small to a rather large value for some range of $\mu^*$ around
 $\mu^*=-3/4 $ bear some similarity to the MF phase diagram. However, due to the much shorter correlation 
length than for $J^*=3$, we can conclude that the  weak periodic order (small amplitude of density oscillations) 
found in MF for relatively large $T^*$  does not resemble an ordered phase in 1d. 

\section{MC simulations in canonical ensemble}

In this section we present results of Monte Carlo simulations for the heat capacity.
The simulations were carried out in the canonical ensemble. The basic step in the sampling
is made as follows: Given the current configuration of the system one chooses at random with equal probability one
of the occupied positions,  $x$ (with $\hat\rho(x)=1$), and one of the empty positions $x'$ (with $\hat\rho(x')=0$), 
the trial configuration is then constructed
by swapping the states between the positions $x$ and $x'$.    Considering the energies of the current and the trial
 configurations one
applies the Metropolis criterion \cite{landau-binder}
to decide whether the trial configuration is accepted as the new configuration of the system
or not.
The heat capacity per particle, $c_v = (\partial (H/N) / \partial T)_{N,L}$ , is computed using the fluctuation formula:

\begin{equation}
c_v = \frac{1}{N k_B T^2} \left[  \langle H^2\rangle - \langle H \rangle^2 \right],
\end{equation}
where the angular brackets indicate averages on the canonical ensemble.

\begin{figure}[th]
\includegraphics[scale=1]{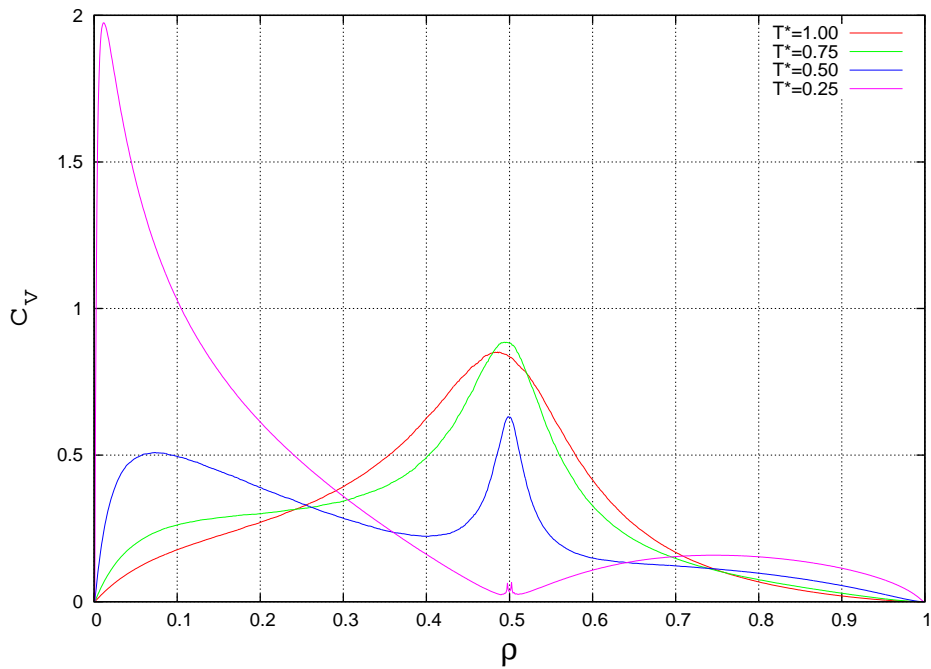}
\caption{The specific heat per particle (in $k_B$ units)  as a function of density (dimensionless) for $J^*=3$ 
 $T^*=0.25,0.5,0.75,1$ with  $L=1200$ . 
}
\label{figCV}
\end{figure}

\begin{figure}[th]
\includegraphics[scale=1]{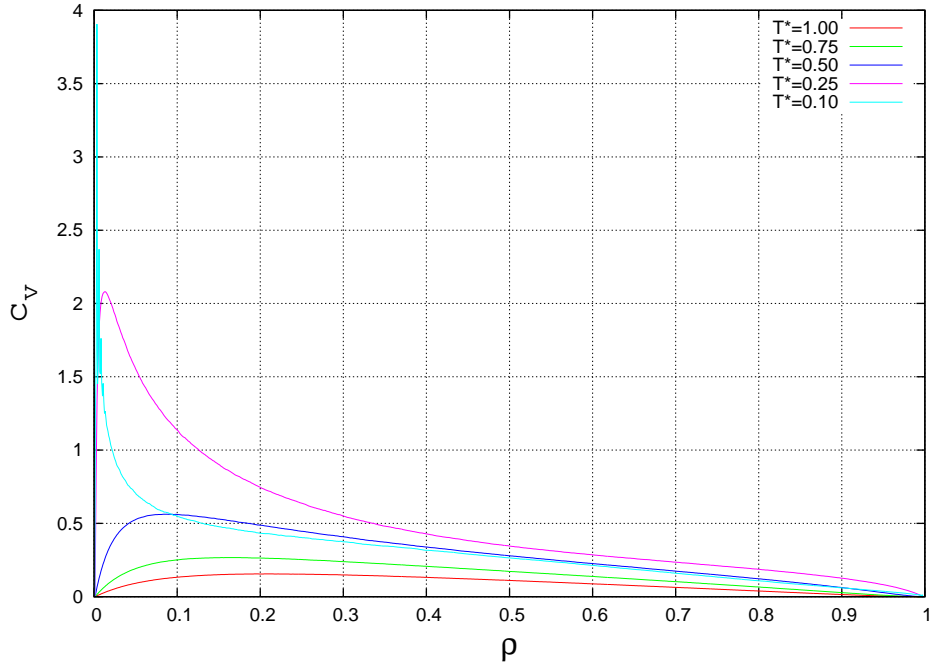}
\caption{The specific heat  (in $k_B$ units) as a function of density  (dimensionless) for $J^*=1/4$ 
and $T^*=0.1,0.25,0.5,0.75,1$ with $L=840$.
}
\label{figCV025}
\end{figure}

For $J^*=3$ the specific heat is shown in Fig.\ref{figCV}.
The results
for $J^*=0.25$ are shown in Fig. \ref{figCV025}.
For both values of $J^*$  we can observe the presence of a peak at low density. On cooling the system the
height of this peak increases, and the density where it appears is shifted to lower values.
This maximum in the heat capacity at low temperature can be explained as an effect of the
equilibrium between isolated particles and clusters of several particles. Given the
Hamiltonian of the model, these clusters are likely to be triples in the case of $J^*=3$.
At low temperature and low density the loss of entropy due to the formation of clusters is
compensated by the energetic effect due to the attractive interaction between the nearest neighbors.

There are, however, significant differences between the heat capacities curves for $J^*=0.25$ and 
$J^*=3.0$, especially for $\rho\simeq 1/2$. 
For $J^*=3.0$,  at low temperature, we can observe a basin around $\rho=1/2$, and a narrow
peak centered also at $\rho=1/2$. Focusing in the region $\rho \le 1/2$, the ground state configurations
are formed by triples of occupied positions. Each triple of ocupied cells is  separated at least by three empty positions from another triple.
Since the triple-triple interaction is repulsive at short distances, the system does not show any trend
to exhibit a pseudo phase separation to form large regions of occupied and empty positions, and therefore the
small energy fluctuations lead to small values of the heat capacity. Notice, however that for $J^*< 1/3$ and low 
temperature the dominant attractive interactions lead to a condensation of particles in large clusters of occupied cells. In this case neither the basin at low $T^*$ nor the peak near $\rho=1/2$ is  present. 

The peak of $c_v(\rho)$ at $\rho \simeq  1/2$ for $J^*=3$ can be interpreted as a signature of  a pseudo-phase
transition between an ordered (or  quasi-ordered) phase (periodic phase with $l=6$ for $\rho=1/2$) and a high temperature 
disordered phase.
Notice that as one approaches $\rho=1/2$ the degeneracy of the ground state reduces sharply, then
we can describe this peak as produced by the competition between the ground state (with
very low entropy values when $\rho \rightarrow 1/2$) and
{\it disordered} states (with higher values of energy and entropy). This ordered {\it pseudo}-phase
lies between the fluid of small droplets ($\rho < 1/2$, $T \rightarrow 0$) and the fluid
of bubbles $(\rho > 1/2$, $T\rightarrow 0$).
In spite of the lack of real phase transitions for one-dimensional models with short range interactions,
the periodic pseudo phase with density $\rho=1/2$ resembles to some extent the low-density crystalline
phases that appear in core-softened models \cite{almarza:09:0} 

 We conclude that measurements of
 $c_v$  in systems with 
competing interactions can give information on the formation and properties of clusters for very small densities, and on formation of phases (or {\it pseudo}-phases) with periodically ordered clusters for higher densities.

\section{summary and discussion }

We have developed a generic model for self-assembly in systems with competing interactions. 
The 1d version of the model was solved in MF approximation and exactly in the grand canonical ensemble for the whole range 
of the repulsion to attraction ratio $J^*$. In addition, MC simulations have been performed in the canonical 
ensemble. Previously lattice models with competing interactions were considered in the context of magnetic
 systems \cite{fisher:80:0,bak:76:0,selke:92:0}. The studies, however, did not focus on the role of the external magnetic
 field that in the
 fluid version of the lattice models  corresponds to the chemical potential $\mu^*$. In the context of fluids the
 chemical potential plays a crucial role and allows to obtain the EOS and structure for dense and diluted fluids.
 We have found interesting and counterintuitive results even in 1d case.

The ground state ($T^*=0$) shows a sequence of phases fluid (gas) - periodic - fluid (liquid) for increasing $\mu^*$ 
when $J^*>1/3$. 
This behavior agrees with the reentrant melting observed experimentaly in several systems \cite{elmasri:12:0,royall:06:0}.
For weaker repulsions only the two fluid phases are present for $T^*=0$. A peculiar property of the MF solutions is the 
existence of the periodic phase for a range of $\mu^*$ that is broader for intermediate temperatures than at $T^*=0$.
 In the case of $1/9<J^*<1/3$ th periodic phase appears for some range of $T^*$, even though it is absent for $T^*=0$. 
Usually, the increase of $T^*$ leads to less ordered structures, and the MF result is counterintuitive. In particular,
 the MF phase diagrams show that for decreasing temperatures at constant $\mu^*$ a sequence of phases fluid-periodic-fluid 
appears, i.e. we find reentrant melting. The high- $T^*$ periodic phase is characterized by a small amplitude of density 
oscillations and a period that is incommensurate with the lattice. 

There are no phase transitions in 1d systems. In order to find if the MF phase behavior is associated with a qualitative
 change of mechanical and structural properties, we have analyzed the  exact results for the EOS and correlation function.
 We have found that for $J^*>1/3$ the shapes of the EOS isotherms and the very large correlation length indicate pseudo phase
 transitions to the periodic phase. On the other hand, the high-$T^*$ weakly ordered phase found in MF is only reflected in
 the qualitative change of properties of the correlation function. For $\mu^*$ that in the ground state corresponds to the
 gas or liquid, we observe that the correlation length assumes a maximum for some finite $T^*$, indicating {\it increasing 
order for increasing $T^*$}. Moreover, for such $T^*$ the amplitude of the correlation function changes from a very small 
value to a much larger value for a narrow range of $\mu^*$, and stays large for the range of $\mu^*$ similar to the stability 
region of the periodic phase found in MF. The most amazing behavior shows the correlation function for $1/9<J^*<1/3$. It 
crosses over from a monotonic decay at low $T^*$ to an oscillatory decay for higher $T^*$ that is, however, much smaller
 than the temperature correponding to the appearance of the periodic phase in MF. The unusual appearence of the  periodic
 short-range order  at intermediate temperatures that for weak repulsion {\it is neither energetically nor entropically } 
favored is the most 
surprising and interesting {\it exact} result of this work. It means that for competing interactions the inhomogeneities 
may appear as a compromise between the macroscopic phase separation that is energetically favorable but entropically
 unfavorable, and the disordered structure that is favorable entropically and unfavorable energetically. 

{\bf Acknowledgment}

We thank E. Lomba and W. T. G\'o\'zd\'z for discussions.
A part of this work  was realized within the International PhD Projects
Programme of the Foundation for Polish Science, cofinanced from
European Regional Development Fund within
Innovative Economy Operational Programme "Grants for innovation". Partial support by the NCN grant is also acknowledged.
NGA gratefully acknowledges financial support from the Direcci\'on
General de Investigaci\'on Cient\'{\i}fica y T\'ecnica under Grant No.
FIS2010-15502, from
the Direcci\'on General de Universidades e Investigaci\'on de la Comunidad de Madrid under Grant
No. S2009/ESP-1691 and Program MODELICO-CM.

\end{document}